\documentclass[twocolumn]{aastex631}

\usepackage{graphicx,textcomp,fancyhdr,hyperref,xcolor,fontawesome}
\definecolor{linkcolor}{rgb}{0.1216,0.4667,0.7059}
\usepackage[caption=false]{subfig}
\usepackage{float}
\usepackage{footnote}
\usepackage[noabbrev]{cleveref}
\usepackage{gensymb}
\usepackage{textcomp}

\shorttitle{exoEarth Volcanism}
\shortauthors{Colby M. Ostberg et al.}

\begin{document}

\title{The Prospect of Detecting Volcanic Signatures on an ExoEarth Using Direct Imaging}

\author[0000-0002-7084-0529]{Colby M. Ostberg}
\affiliation{Department of Earth and Planetary Sciences, University of California, Riverside, CA 92521, USA}
\affiliation{NASA Goddard Space Flight Center, 8800 Greenbelt Road, Greenbelt, MD 20771, USA}
\email{costb001@ucr.edu}

\author[0000-0003-1149-7385]{Scott D. Guzewich}
\affiliation{NASA Goddard Space Flight Center, 8800 Greenbelt Road, Greenbelt, MD 20771, USA}
\affiliation{Sellers Exoplanet Environments Collaboration, NASA Goddard Space Flight Center, 8800 Greenbelt Road, Greenbelt, MD 20771, USA}

\author[0000-0002-7084-0529]{Stephen R. Kane}
\affiliation{Department of Earth and Planetary Sciences, University of California, Riverside, CA 92521, USA}

\author[0000-0002-0000-199X]{Erika Kohler}
\affiliation{NASA Goddard Space Flight Center, 8800 Greenbelt Road, Greenbelt, MD 20771, USA}

\author[0000-0002-5487-2598]{Luke D. Oman}
\affiliation{NASA Goddard Space Flight Center, 8800 Greenbelt Road, Greenbelt, MD 20771, USA}
\affiliation{Sellers Exoplanet Environments Collaboration, NASA Goddard Space Flight Center, 8800 Greenbelt Road, Greenbelt, MD 20771, USA}

\author[0000-0002-5967-9631]{Thomas J. Fauchez}
\affiliation{Integrated Space Science and Technology Institute, Department of Physics, American University, Washington DC}
\affiliation{NASA Goddard Space Flight Center, 8800 Greenbelt Road, Greenbelt, MD 20771, USA}
\affiliation{Sellers Exoplanet Environments Collaboration, NASA Goddard Space Flight Center, 8800 Greenbelt Road, Greenbelt, MD 20771, USA}

\author[0000-0002-5893-2471]{Ravi K. Kopparapu}
\affiliation{NASA Goddard Space Flight Center, 8800 Greenbelt Road, Greenbelt, MD 20771, USA}
\affiliation{Sellers Exoplanet Environments Collaboration, NASA Goddard Space Flight Center, 8800 Greenbelt Road, Greenbelt, MD 20771, USA}

\author[0000-0001-8138-7582]{Jacob Richardson}
\affiliation{NASA Goddard Space Flight Center, 8800 Greenbelt Road, Greenbelt, MD 20771, USA}
\affiliation{Sellers Exoplanet Environments Collaboration, NASA Goddard Space Flight Center, 8800 Greenbelt Road, Greenbelt, MD 20771, USA}

\author[0000-0003-3266-9772]{Patrick Whelley}
\affiliation{University of Maryland, Department of Astronomy, College Park, MD 20742, USA}
\affiliation{NASA Goddard Space Flight Center, 8800 Greenbelt Road, Greenbelt, MD 20771, USA}
\affiliation{Sellers Exoplanet Environments Collaboration, NASA Goddard Space Flight Center, 8800 Greenbelt Road, Greenbelt, MD 20771, USA}


\begin{abstract}
The James Webb Space Telescope (JWST) has provided the first opportunity to study the atmospheres of terrestrial exoplanets and estimate their surface conditions. Earth-sized planets around Sun-like stars are currently inaccessible with JWST however, and will have to be observed using the next generation of telescopes with direct imaging capabilities. Detecting active volcanism on an Earth-like planet would be particularly valuable as it would provide insight into its interior, and provide context for the commonality of the interior states of Earth and Venus. In this work we used a climate model to simulate four exoEarths over eight years with ongoing large igneous province eruptions with outputs ranging from 1.8--60 Gt of sulfur dioxide. The atmospheric data from the simulations were used to model direct imaging observations between 0.2--2.0 $\mu$m, producing reflectance spectra for every month of each exoEarth simulation. We calculated the amount of observation time required to detect each of the major absorption features in the spectra, and identified the most prominent effects that volcanism had on the reflectance spectra. These effects include changes in the size of the O$_3$, O$_2$, and H$_2$O absorption features, and changes in the slope of the spectrum. Of these changes, we conclude that  the most detectable and least ambiguous evidence of volcanism are changes in both O$_3$ absorption and the slope of the spectrum.
\end{abstract}



\section{Introduction}
\label{sec:intro}

The atmospheres of terrestrial exoplanets have become increasingly accessible through the use of both transmission and emission spectroscopy \citep{seager2010,madhusudhan2019}, particularly in the era of the James Webb Space Telescope \citep[JWST;][]{beichman2014b,greene2016,bean2018,fortenbach2020}. Emission spectroscopy and thermal infrared phase curves have been used to determine whether exoplanets have atmospheres \citep[e.g.;][]{kriedberg2019,kane2020d,greene2023thermal}, while transmission spectroscopy is capable of identifying the molecular species in exoplanet atmospheres \citep[e.g.][]{sing2011hubble,ridden2023high,tinetti2010probing,fraine2014water}. These techniques provide vital information that can be input into three-dimensional (3-D) general circulation models (GCMs) to estimate the potential climate states of exoplanets \citep[e.g.][]{wolf2017a,wolf2017b,fauchez2021trappist,turbet2016habitability,turbet2018modeling}. Refining estimates of exoplanet climates will require understanding the states of both the atmosphere and the interior of a planet, since geological activity has been a crucial component for Earth maintaining habitable conditions throughout its history.

Determining the geological properties of an exoplanet by studying its surface is a challenging task with current technologies. Therefore, inferring the state of exoplanet interiors will largely rely on indirect techniques, such as through the study of planetary atmospheres \citep{kislyakova2017,guenther2020,harnett2020,quick2020,noack2021}. Detection of volcanic compounds such as methane (CH$_4$), carbon dioxide (CO$_2$), and sulfur dioxide (SO$_2$) would provide insight into potential volcanic activity occuring on an exoplanet \citep[e.g.][]{misra2015transient,kaltenegger2010detecting,fortin2022volcanic,hu2013photochemistry,loftus2019sulfate}. Detection of SO$_2$ in an exoEarth atmosphere would be a particularly strong indication of ongoing volcanism due to its short chemical lifetime in an Earth-like atmosphere. Atmospheric CO$_2$ has a much longer lifetime than SO$_2$ and is evidence of outgassing, however it would be difficult to constrain whether the outgassing occurred recently or in the past. CH$_4$ also has a short lifetime in Earth's atmosphere, but would require either substantial volcanism or living organisms in order for detectable amounts of CH$_4$ to be sustained \citep[e.g.][]{thompson2022case,krissansen2018disequilibrium,krissansen2022understanding}. JWST has the capability to probe the atmospheres of terrestrial planets, but is limited to those that orbit cooler, smaller stars such as M-dwarfs. Observing the atmosphere of an Earth-like planet around a Sun-like star will require future direct imaging missions like the Habitable Worlds Observatory (HWO), which is based on the HabEx \citep{gaudi2020habitable} and LUVOIR \citep{luvoir2019luvoir} mission concepts, and the Large Interferometer For Exoplanets (LIFE), which would be sensitive enough to directly image planets around M-dwarfs \citep{quanz2022large}.

In this work, we investigate the potential of detecting evidence of volcanism in the reflectance spectra of an exoEarth. This involves using a 3-D GCM to simulate volcanic eruptions of varying size on an exoEarth, identifying how volcanism affects the reflectance spectra of each eruption, and simulating observations with a LUVOIR-like telescope. In Section~\ref{sec:Methods}, we describe the 3-D GCM used to model eruptions, and how we simulated the reflectance spectra and observations of the exoEarth. Section~\ref{sec:results} describes the results of the S/N analysis and the maximum and minimum observation time needed to detect individual features. In Section~\ref{sec:discussion}, we discuss the features and bandpasses which should be prioritized in future observations, and the caveats of the study. Concluding remarks and future work that is needed are discussed in Section~\ref{sec:conclusions}.


\section{Methods}
\label{sec:Methods}

\subsection{GEOSCCM Global Climate Model}
\label{sec:geosccmmodel}

The Goddard Earth Observing System Chemistry Climate Model (GEOSCCM) simulates Earth's modern climate using coupled atmospheric general circulation and dynamic ocean models \citep{rienecker2008geos,nielsen2017chemical,aquila2021impacts}. It additionally uses the Global Modeling Initiative (GMI) stratosphere-troposphere chemical routine \citep{strahan2007observationally,duncan2007model} that integrates a bulk aerosol module (Goddard Chemistry, Aerosol, Radiation, and Transport (GOCART)) \citep{colarco2010online}. In combination, this allows GEOSCCM to self-consistently simulate climate, cloud, chemistry, and aerosol physics. Simulations were run at 1$^{\circ}$ x 1$^{\circ}$ horizontal resolution in both the ocean and atmosphere, with 72 vertical layers in the atmosphere extending to ~80 km altitude and 50 ocean layers to a depth of 4.5 km. The model's initial and boundary conditions are identical to the modern pre-industrial Earth and Sun with fixed CO$_2$ concentrations of 280 ppm.    

Our simulations modeled the climate impact of large igneous province (LIP) (also known as "flood basalt") volcanism \citep{bond2021global,bryan2010largest,courtillot2003ages,self2006volatile}. Specifically, the Columbia River flood basalt eruption, which is geologically the most recent (~15-17 Ma) and smallest known such eruption in terrestrial history \citep{mckay2014estimating,kasbohm2018rapid}. Flood basalt volcanism is believed to occur on every other terrestrial world in the solar system \citep{o2000flood,lancaster1995great,head2011flood,jaeger2010emplacement}. The model simulates such an eruption by injecting SO$_2$ in both the near-surface atmosphere and the upper troposphere-lower stratosphere over the course of 4 years at two model grid points in eastern Oregon and Washington states, USA. 

We simulated 6 eruptions which vary from 1.875--60 Gt of emitted SO$_2$, and a baseline case with no SO$_2$ release. The SO$_2$ output of the simulated eruptions are similar to that of the 1815 Tambora eruption \citep{stothers1984great}, and the Toba volcano in Sumatra \citep{oppenheimer2002limited}. For reference, the yearly global output of SO$_2$ from volcanoes on Earth is estimated to be 0.01 Gt \citep{stoiber1973sulfur}. After the SO$_2$ is emitted in the simulations, the SO$_2$ can be oxidized by O$_2$ and hydroxide (OH), and then combined with water (H$_2$O) to form sulfuric acid aerosols (H$_2$SO$_4$; hereafter referred to as volcanic aerosol). This can occur in the model at any pressure where the chemistry is suitable.

Additional details of the volcanic eruption scenario and GEOSCCM as it was used here are provided by \citet{guzewich2022volcanic}. In brief, explosive eruptions (placing SO$_2$ into the upper troposphere and lower stratosphere) occur once every 3 months for the first 4 simulated years with near-surface effusive eruptions continuing throughout those 4 years. The eruptions stop after the first 4 years, and the simulations were ran for an additional 4 years to examine any lasting changes to the climate system.

Each of the 5 (4 volcanic and 1 baseline) simulations are run for a total of 8 simulated years with variables output as monthly averages. For our purposes of evaluating the planetary reflected light spectrum, monthly average values better represent the expected spectral changes caused by aerosols, clouds, and variable gas abundances due to the eruption and the subsequent climate impacts rather than changes caused by weather on the timescales of hours to days.

\begin{figure}
\centering 

\includegraphics[width=\columnwidth]{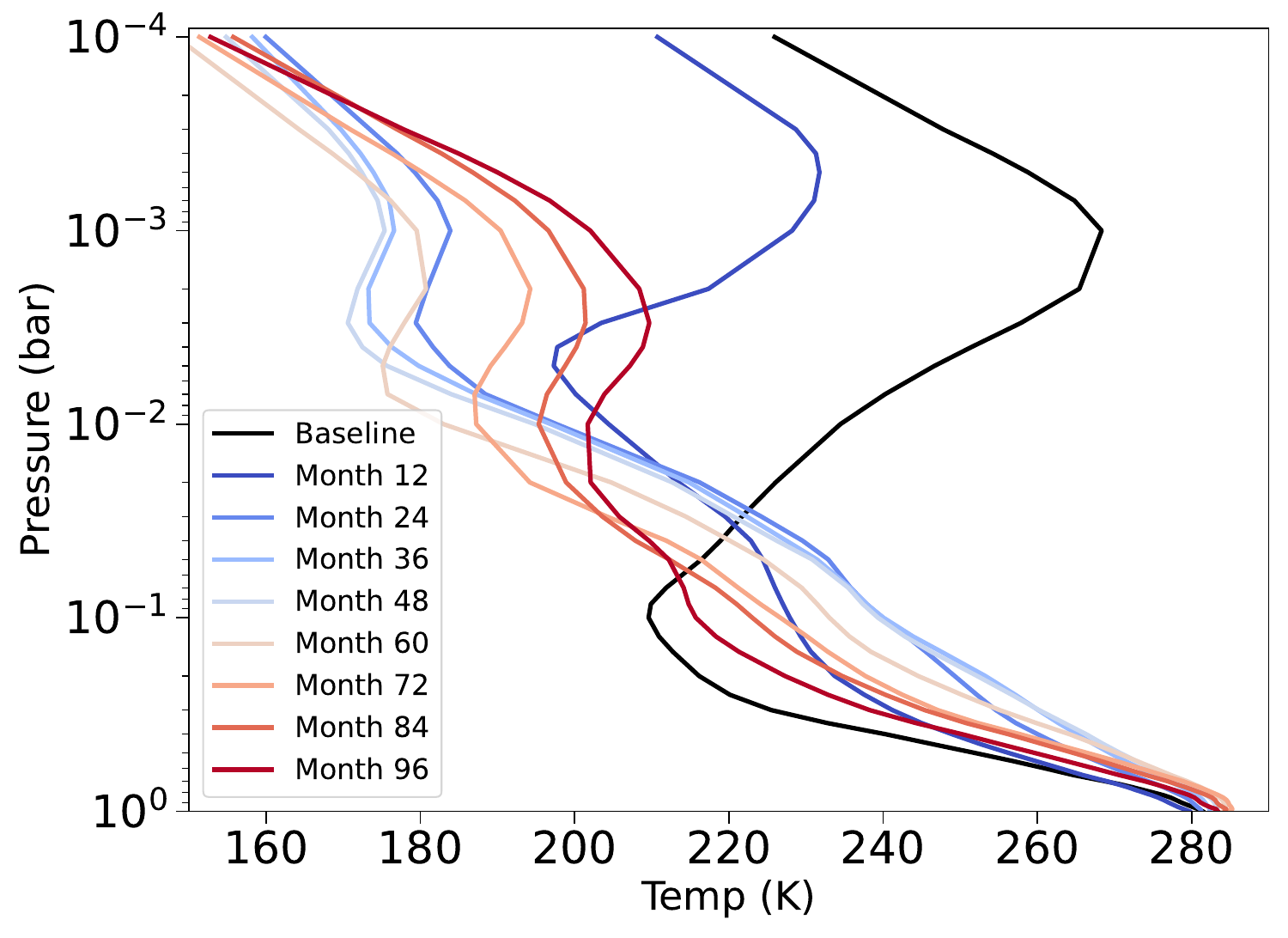}%

\includegraphics[width=\columnwidth] {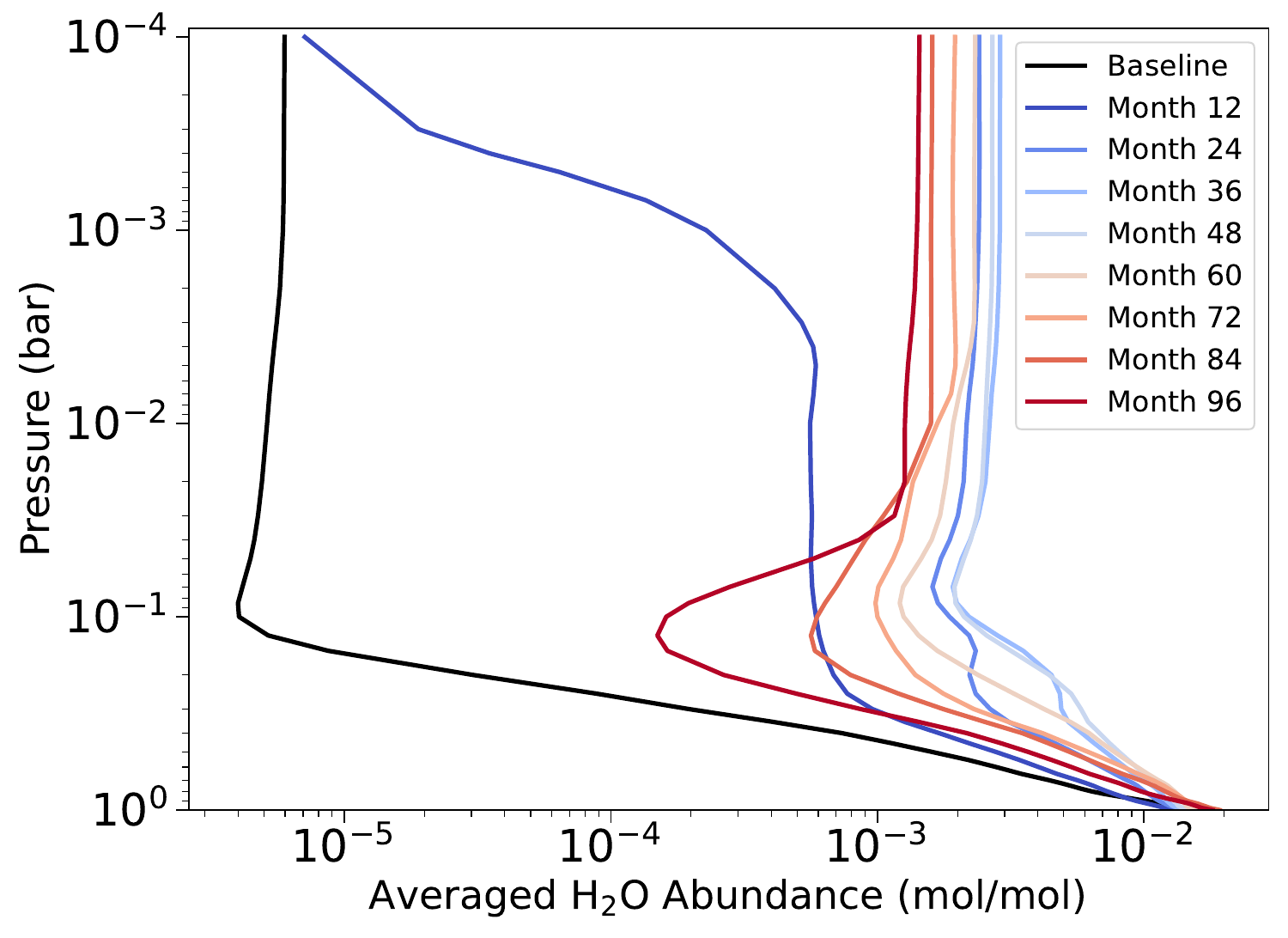}%

\includegraphics[width = \columnwidth]{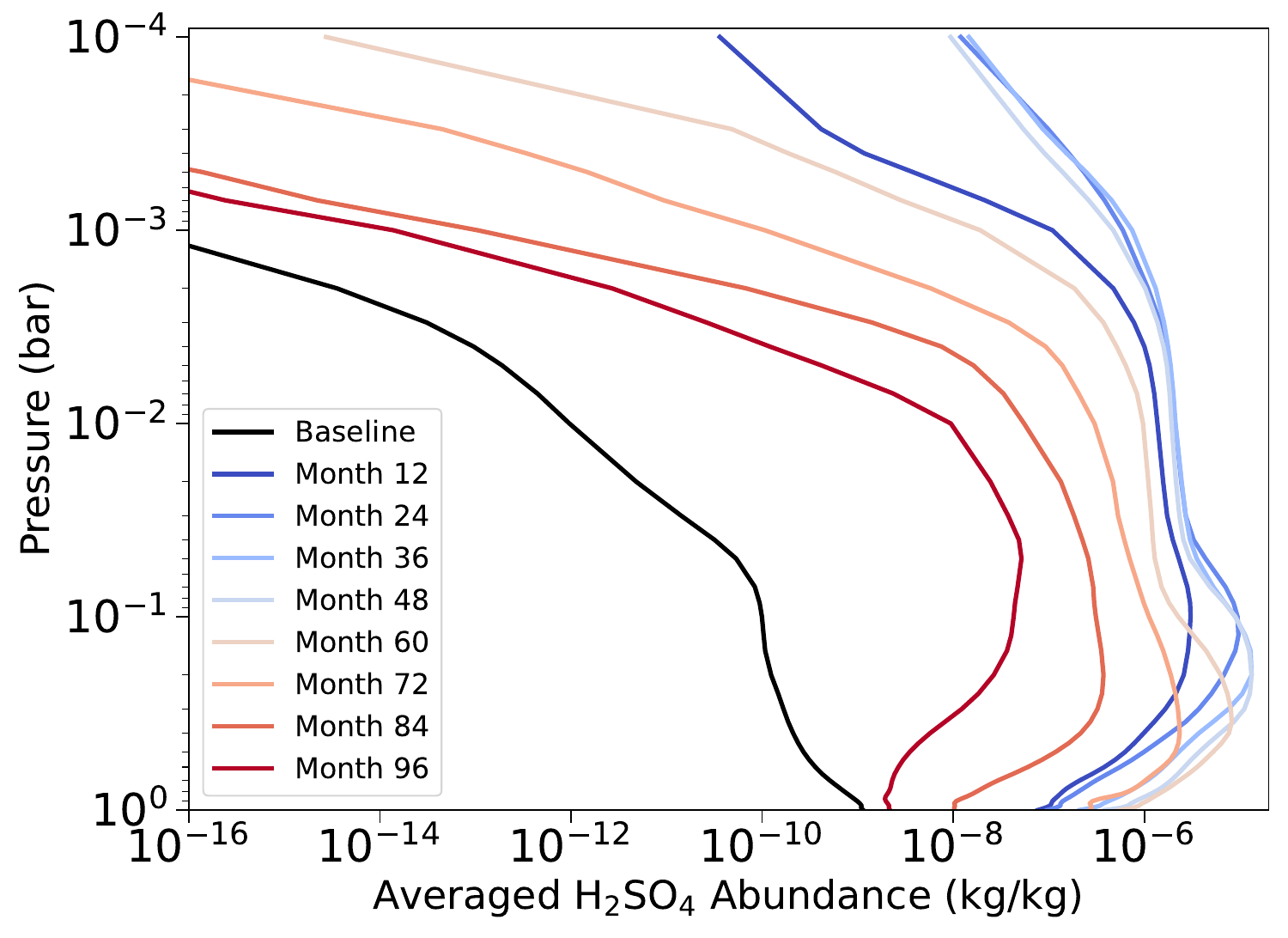}

\caption{The globally averaged TP profiles (upper panel), H$_2$O vapor abundance profiles (middle panel), and volcanic aerosol abundance profiles (lower panel) of the 30 Gt eruption exoEarth for every 12 months of the simulation. The corresponding profile of the baseline simulation is shown for reference.  
\label{fig:TPs}}
\end{figure}

The globally averaged temperature-pressure profiles for every 12 months of the 30 Gt eruption simulation and the single year of the baseline simulation are shown in the upper panel of Figure \ref{fig:TPs}. The volcanic aerosols heat up the upper troposphere and lower stratosphere which remove the tropopause seen in the baseline TP profile. The eruptions also lead to depletion of O$_3$, which causes the upper atmosphere to be much cooler in the eruption simulation than that of the baseline simulation. After the eruptions cease, the tropopause inversion begins to return near the end of the eruption simulation as the volcanic aerosols are slowly removed and O$_3$ is replenished. The eruptions also move large amount of H$_2$O vapor into the troposphere and stratosphere, as shown in the middle panel of Figure \ref{fig:TPs}, which also contributes to the removal of the tropopause. The lower panel of Figure \ref{fig:TPs} illustrates the large increase in volcanic aerosols during the first 4 years of the simulation, and the slow decrease in volcanic aerosols throughout the last 4 years of the simulation.

\setlength{\belowcaptionskip}{2pt}


\subsection{GlobES and PSG}
\label{sec:globespsg}

In order to model the reflected light spectra of the GEOSCCM exoEarths, we input the monthly averaged outputs into the Global Emission Spectra (GlobES, \url{https://psg.gsfc.nasa.gov/apps/globes.php}) application, which is part of the Planetary Spectrum Generator (PSG, \url{https://psg.gsfc.nasa.gov}) radiative transfer suite. GlobES uses the 3-D TP and chemical abundance data from GCMs which allows it to incorporate the effects of an inhomogenous atmosphere and surface on reflected light spectra. 

We modelled a timeline of reflectance spectra for each of the 4 volcanically active exoEarths and the baseline exoEarth. All exoEarths were defined to have Earth's radius and mass with a circular orbit at 1 AU around a Sun-like star that is 10 parsecs away. The inclination of the system was defined to be edge-on and the planet had a phase angle of 90$^{\circ}$ in reference to the observer. The atmosphere of each exoEarth was defined using their monthly averaged atmospheres. Each volcanic exoEarth was simulated for 96 months, giving them 96 reflectance spectra each. The baseline exoEarth was only simulated for 12 months so it has 12 reflectance spectra.

We assumed the exoEarth is observed by a hypothetical LUVOIR-B telescope with a 6-meter mirror and attached coronagraph. Observational noise was simulated for the ultraviolet (UV), visible (VIS), and near-infrared (NIR) instruments which have bandpasses of 0.2 -- 0.515, 0.515 -- 1.0, and 1.01 -- 2.0 $\mu$m, respectively. All simulated observations consisted of a single 1-hour exposure. The uncertainty of longer observations were extrapolated using a scaling relationship that is explained in more detail in Section \ref{sec:CalcSN}. The values for instrument-related input parameters were the same as those used in \cite{checlair2021probing}, which were chosen using the LUVOIR final report \citep{luvoir2019luvoir}. The instrument inputs were also similar to that of \cite{kopparapu2021nitrogen}, who modelled observations of LUVOIR-A. 

\begin{figure*}
  \includegraphics[width = \textwidth]{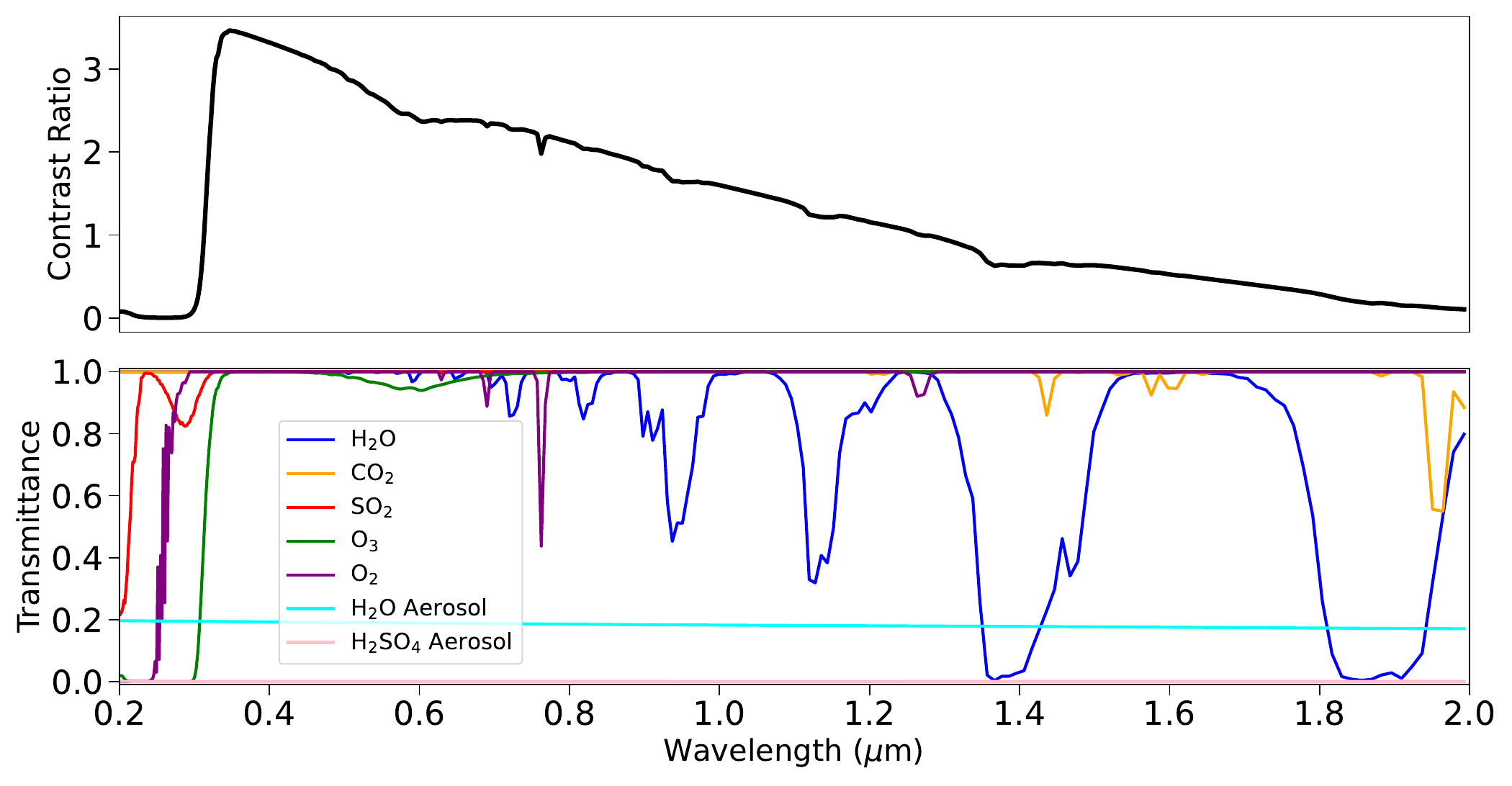}
  \caption{The reflectance spectra of the 30 Gt eruption exoEarth in Month 11 of the simulation (upper panel) and the transmittances of the major molecular species in its atmosphere (lower panel). The most prominent absorption feature in the UV is caused by O$_3$. Absorption by SO$_2$ occurs around in the UV around 0.3 $\mu$m, however it is overshadowed by O$_3$ and O$_2$ absorption. The VIS bandpass has a single O$_2$ absorption band, and multiple H$_2$O absorption bands. The NIR bandpass includes three H$_2$O absorption bands at 1.15, 1.4, and 1.85 $\mu$m.}
  \label{fig:208411_Transmittance}
\end{figure*}

The atmospheric parameters used as inputs for GlobES include 3-D TP, molecular abundance, and aerosol abundance profiles. The gas species include H$_2$O vapor, CO$_2$, ozone (O$_3$), nitrous oxide (N$_2$O), SO$_2$, CH$_4$, oxygen (O$_2$), and nitrogen (N$_2$). The aerosol species include volcanic aerosol, H$_2$O aerosol, and water-ice. Figure \ref{fig:208411_Transmittance} shows the reflectance spectra and corresponding molecular transmittance of the 30 Gt eruption exoEarth atmosphere during month 11 of the simulation. The UV bandpass is dominated by O$_3$ absorption, but also contains O$_2$ and SO$_2$ absorption bands. The SO$_2$ absorption around 0.3 $\mu$m is present during the first 4 years of the simulations while volcanism is ongoing, however, it is always overshadowed by the O$_3$ and O$_2$ absorption bands that it overlaps with. The VIS bandpass includes absorption by O$_2$ at 0.75 $\mu$m, smaller H$_2$O features which straddle the O$_2$ feature, and a larger H$_2$O feature at 0.95 $\mu$m. The NIR bandpass has 3 H$_2$O absorption features at 1.15, 1.4, and 1.9 $\mu$m. During this month, H$_2$O and volcanic aerosols are also present which inhibit the size of the H$_2$O absorption features.

\subsection{Calculating S/N of Spectral Features}
\label{sec:CalcSN}

\begin{figure*}[!htbp]
\centering 
\subfloat{%
  \includegraphics[width=0.7\columnwidth]{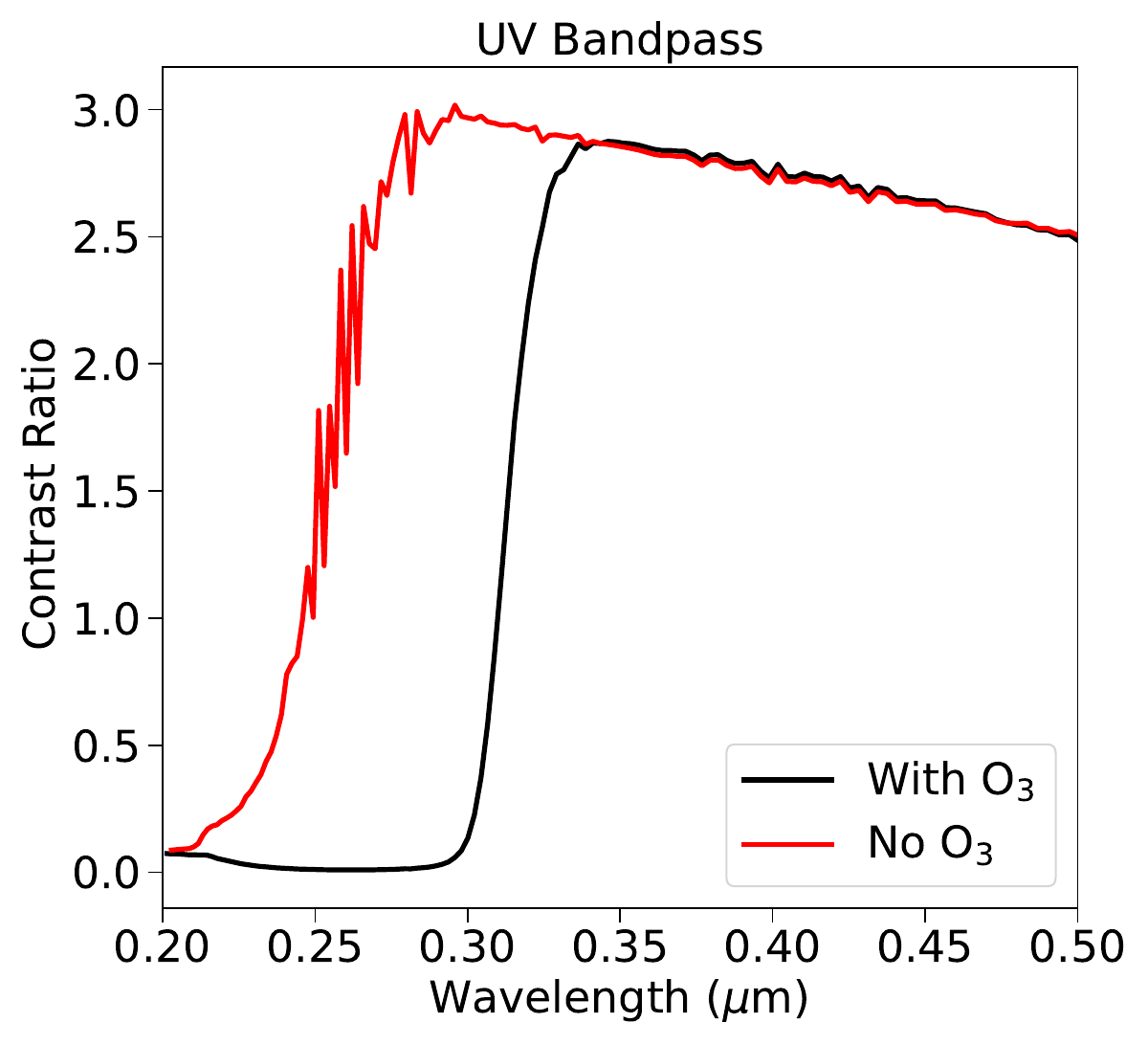}%
}
\subfloat{%
  \includegraphics[width=0.7\columnwidth]{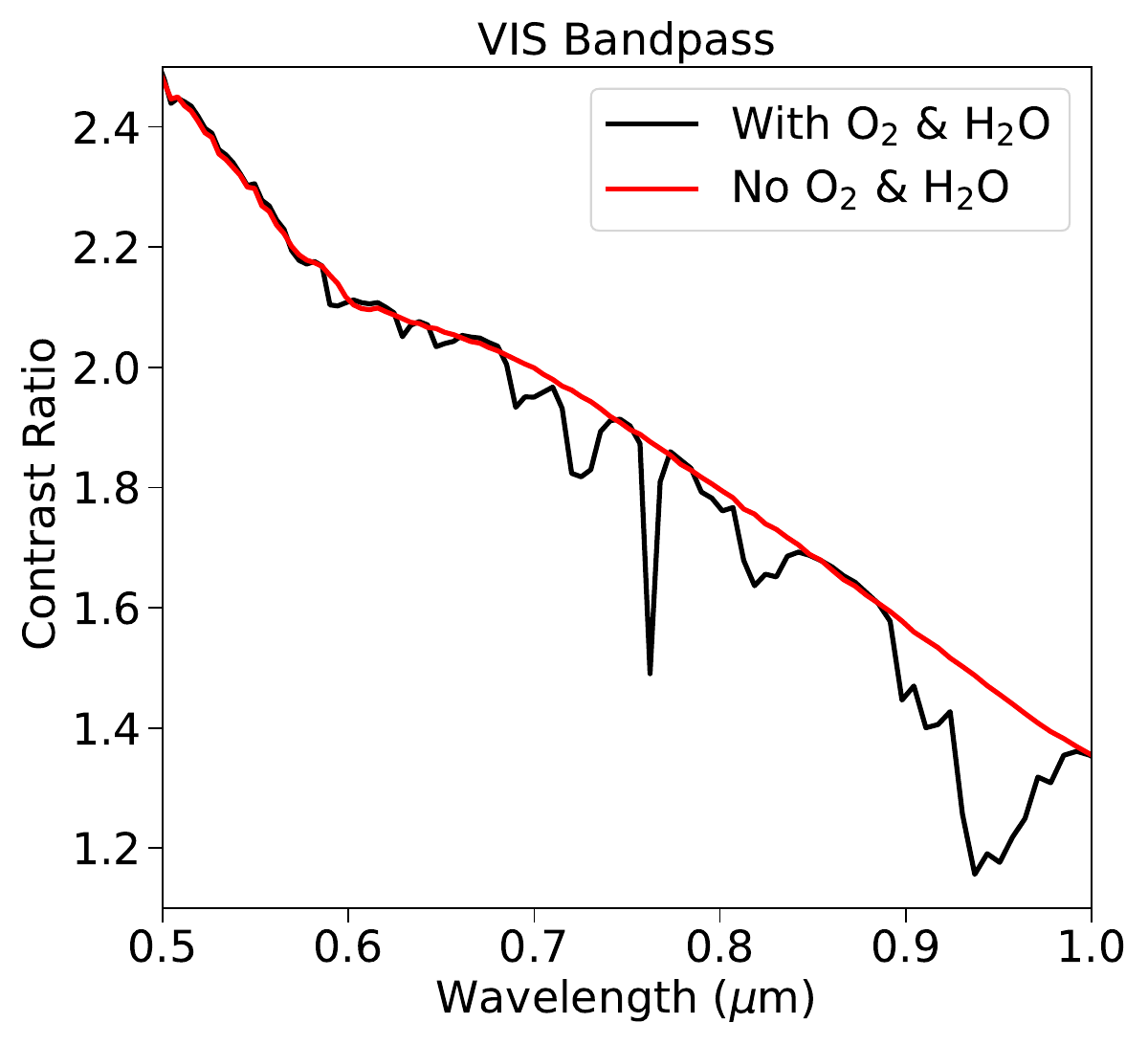}%
}
\subfloat{%
  \includegraphics[width=0.7\columnwidth]{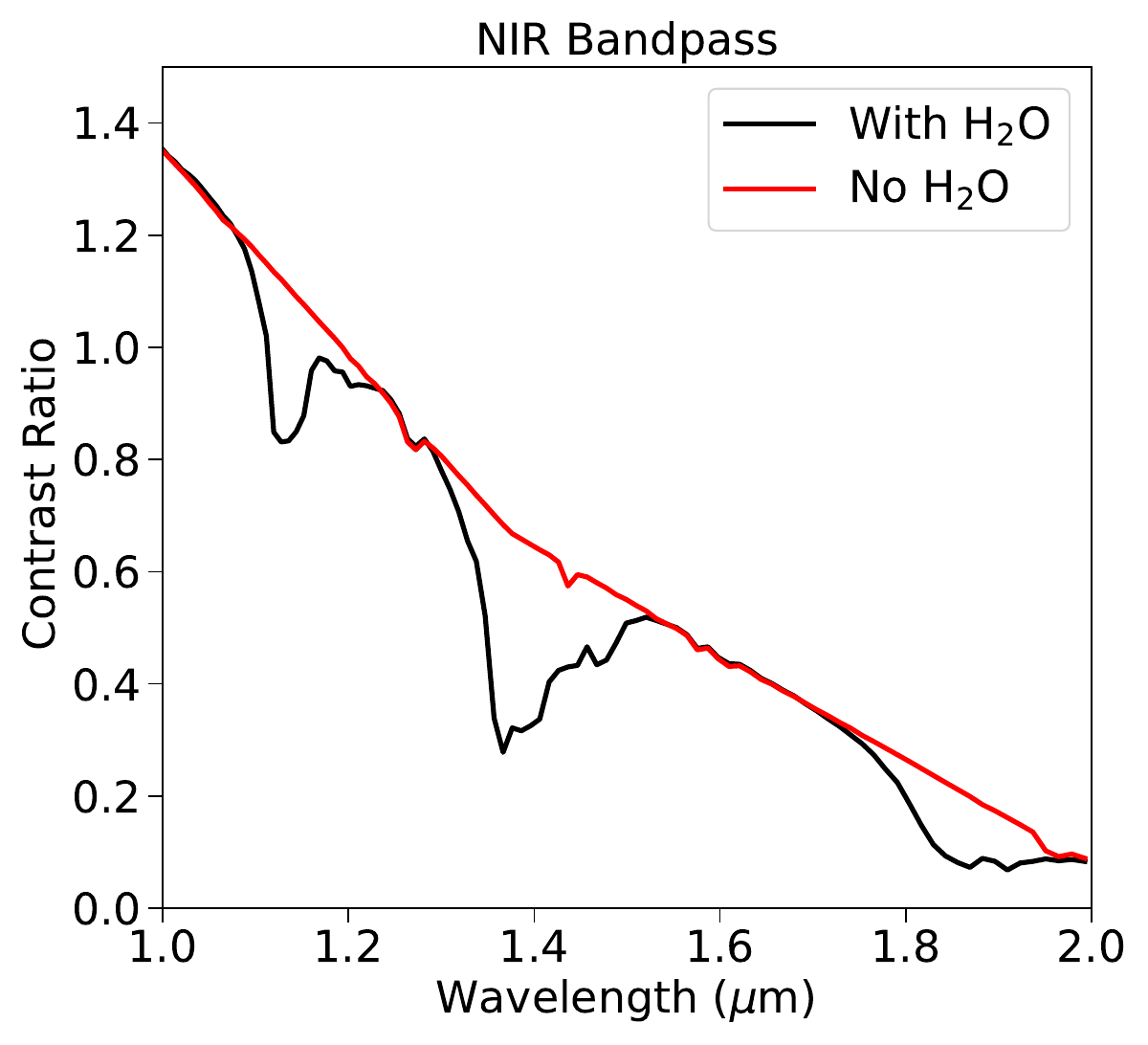}%
}
\caption{The reflectance spectrum of the 15 Gt eruption exoEarth in month 71 of the simulation with and without absorption from O$_3$, H$_2$O, or O$_2$. The left panel shows the UV spectrum without O$_3$ absorption. The middle panel shows the VIS spectrum without O$_2$ and H$_2$O absorption. The right panel shows the NIR spectrum without H$_2$O absorption.
\label{fig:NoAbsorption}}
\end{figure*}

We computed the signal-to-noise ratio (S/N) of the major molecular features found in the UV, VIS, and NIR bandpasses in order to quantify their detectability. We chose to focus on calculating S/N for the O$_3$ feature in the UV bandpass, O$_2$ and H$_2$O features in the VIS bandpass, and the H$_2$O features in the NIR bandpass. The S/N calculated for the H$_2$O features in the VIS and NIR bandpasses is the combined S/N of all H$_2$O features in the given bandpass. To determine the S/N of each feature, we used Equation \ref{SNEquation} which is a $\chi^2$ approach used previously by \citet{lustigyaeger2019a}:
\begin{equation}
    S/N = \sqrt{ \sum_{i}^{N_{\lambda_i}} \left ( \frac{y_i - y_{cont}}{\sigma_i} \right )^2 }
\label{SNEquation}
\end{equation}
In Equation \ref{SNEquation}, $y_i$ is the i'th y value of the modelled spectrum within the given instrument's bandpass, $\sigma_i$ is the corresponding uncertainty of the simulated data, and $y_{cont}$ is the continuum of the spectrum which we defined to be the same spectrum but without absorption from the molecule of interest. Figure \ref{fig:NoAbsorption} shows the continua used to calculate S/N of the respective features in the UV, VIS, and NIR bandpasses. The S/N of each feature was first calculated for 1 hour of observation, then a scaling relationship was used to extrapolate the S/N of longer observations in 1 hour intervals assuming a photon noise-limited scenario. The S/N values estimated by the scaling relationship were compared to the S/N values from fully computing the noise of the data. Both methods yielded very similar S/N values, so we opted to use the scaling relationship for all S/N calculations as it is much faster.

We defined the threshold for detection of a feature to be when the S/N equals or exceeds a value of 5, which is the same threshold that has been used in previous studies \citep[e.g.][]{lustig2019detectability,pidhorodetska2020detectability,felton2022role}. The observation time required to detect individual features were calculated for every monthly averaged spectrum in all eruption cases. The detection of molecules from actual observations will require more complex analysis with retrieval algorithms and model comparisons, therefore the observation times needed for detection reported in this work are to be considered as lower limits.


\section{Results}
\label{sec:results}

\subsection{The Effect of Volcanism on Reflectance Spectra}
\label{sec:spectra}

\begin{figure*}[!htbp]
  \includegraphics[width = \textwidth]{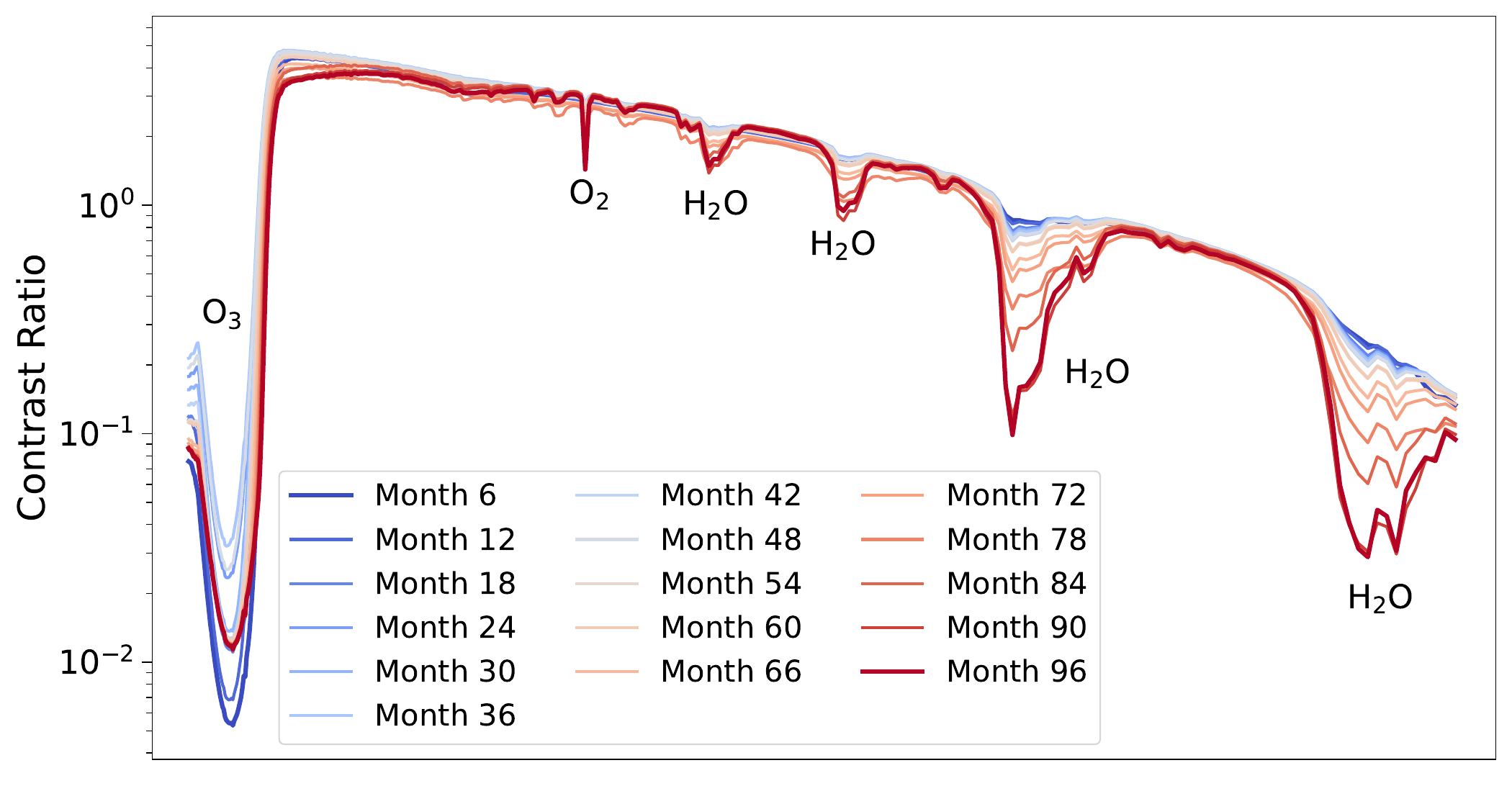}
  \includegraphics[width = \textwidth]{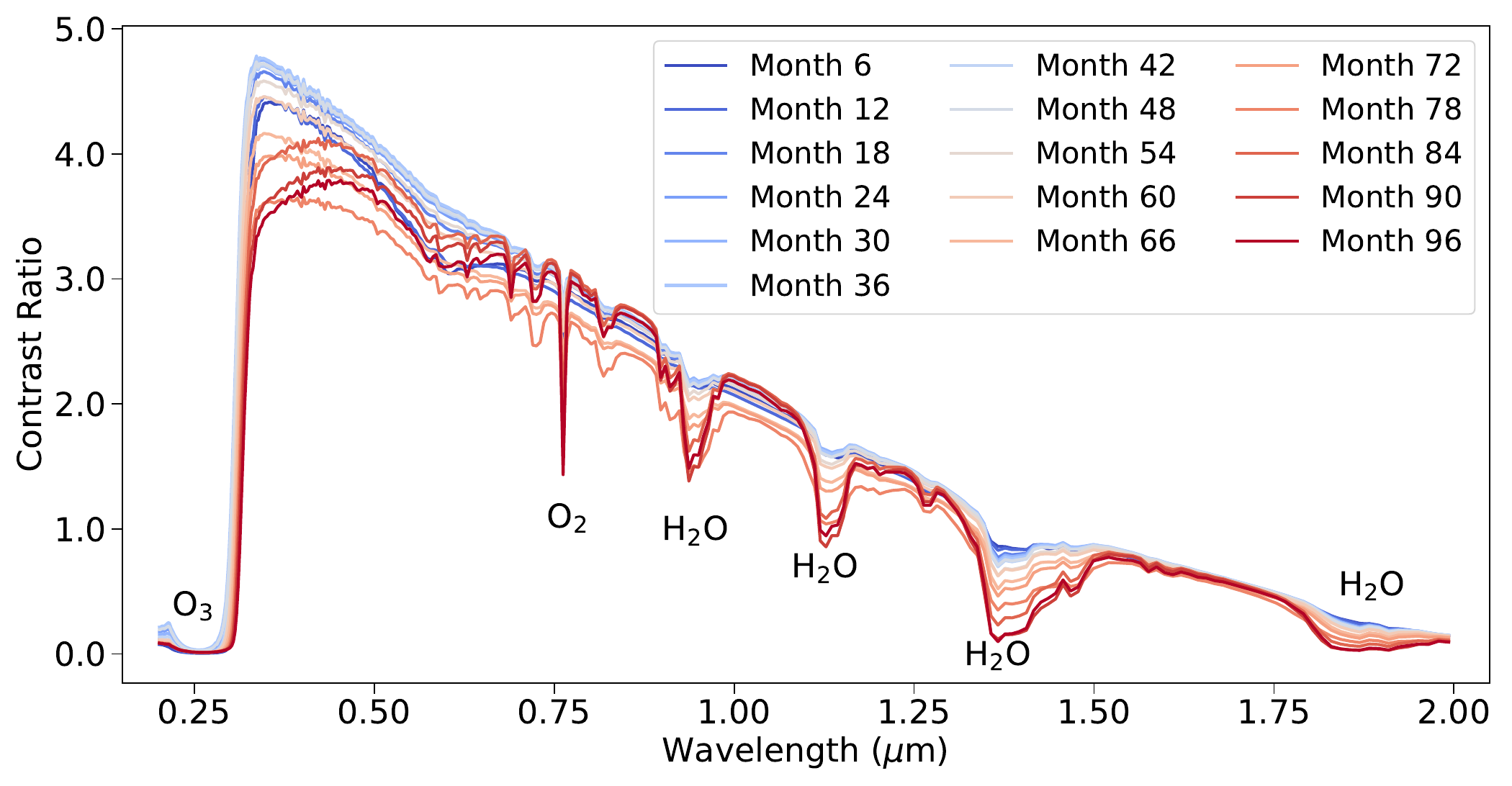}
  \caption{The reflectance spectrum of the 30 Gt eruption exoEarth for every 6 months of the simulation. Absorption features are labelled with their corresponding molecules. Both panels show the same spectra but the upper panel is on a log scale while the lower panel is scaled linearly. There is variation in the size of the O$_3$ absorption feature and slope of the spectra between 0.3--0.6 $\mu$m due to fluctuations in volcanic aerosol abundance throughout the simulation. The change in slope of the spectrum also affects the location of the continuum. The size of the H$_2$O features change throughout the simulation due to the eruptions causing an influx of volcanic aerosols and H$_2$O vapor into the upper atmosphere.}
  \label{fig:6monthplot}
\end{figure*}

Figure \ref{fig:6monthplot} shows the reflectance spectrum of the 30 Gt eruption exoEarth for every 6 months of the simulation in both a log (upper panel) and linear scale (lower panel). The spectra are shown in units of contrast ratio, which is the radiance of the planet divided by the radiance of the star. Since the stars are much brighter than planets, the contrast ratio of a direct imaging observation will be less than 1. However, since we included a coronograph in our simulated observations, which blocks most of the radiance from the star, the contrast ratio of the spectra in Figure \ref{fig:6monthplot} are greater than 1 at certain wavelengths

The dominant absorber in the UV bandpass is O$_3$, which causes the dip at 0.25 $\mu$m (Figure \ref{fig:6monthplot}). The volcanic aerosols produced from the eruptions serve as a catalyst for reactions that deplete O$_3$, and therefore reduces the amount of light absorbed by O$_3$. The maximum amount of volcanic aerosols is reached in month 48, which in turn causes the month 48 spectrum to have the least absorption by O$_3$. The eruptions stop after month 48 in the simulation, which allows the O$_3$ abundance to begin to replenish and the size of the O$_3$ feature to increase. There is also variation in the slope of the spectra throughout the simulation, where a sharp peak forms around 0.4 $\mu$m during the first 48 months of the simulation due to volcanic aerosols increasing scattering in the atmosphere. The peak subsides during the last 48 months of the simulation as chemical reactions remove volcanic aerosols from the atmosphere.

The presence of aerosols also affects the O$_2$ and H$_2$O absorption features in the VIS bandpass. Since O$_2$ abundance was assumed to be constant during the simulation, changes to the O$_2$ feature are only caused by aerosol scattering in the upper atmosphere which conceals the O$_2$ absorption in the lower atmosphere and decreases the size of the feature. Despite the eruptions transporting H$_2$O vapor into the upper atmosphere, the size of the H$_2$O absorption features in both the VIS and NIR bandpasses are stinted for the first 48 months in the simulations because of the presence of volcanic aerosols. Once the eruptions stop, the abundance of volcanic aerosols begins to quickly deplete via chemical reactions, whereas H$_2$O vapor is removed much slower and is able to remain abundant for years after the eruptions. This can be seen in the spectra as the H$_2$O absorption features remain small during the first 48 months of the simulation, but increase in size during the last 48 months of the simulation (Figure \ref{fig:6monthplot}).


\subsection{Detecting Absorption Features}
\label{sec:detectfeatures}

Using Equation \ref{SNEquation}, we determined the amount of observation time required to detect (S/N~$\geq$~5) the absorption features in the reflectance spectra for each month of every exoEarth simulation. Tables \ref{tab:UV_SN}, \ref{tab:VIS_SN}, and \ref{tab:NIR_SN} list the maximum (Max), minimum (Min), and average (Avg) time to detect the major features in each bandpass of the 4 volcanic exoEarth spectra. We considered all 96 spectra for a given exoEarth simulation when determining the Max, Min, and Avg detection time for each feature.

Table \ref{tab:UV_SN} lists the observation time needed to detect the O$_3$ absorption feature in the UV bandpass for the 4 volcanic exoEarths. The O$_3$ feature is unique in that it is easily identifiable even when O$_3$ absorption is decreased. Furthermore, the feature becomes easier to detect when there is less O$_3$ absorption because the peak of the feature extends to a larger contrast ratio, which increases the signal of the feature. As a result, the 15, 30, and 60 Gt eruptions have lower minimum detection times than the 1.8 Gt eruption. For all months in every exoEarth simulation, the O$_3$ feature required no more than 5 hours of observation to be detected, making it the most consistently detectable feature in any bandpass.

\begin{deluxetable}{|c|c|c|c|}
\tablecaption{Observation Hours Required to Detect O$_3$ in the UV \label{tab:UV_SN}}
\tablehead{\colhead{Eruption (Gt)}  & \colhead{O$_3$ Max}  & \colhead{O$_3$ Min}  & \colhead{O$_3$ Med} }
\startdata
1.8 & 4 & 3 & 3 \\
15 & 4 & 2 & 2 \\
30 & 4 & 2 & 2 \\
60 & 5 & 2 & 2 \\
\enddata
\tablenotetext{}{Max, Min, and Med designate the maximum, minimum, and median time needed to detect the associated feature during each simulation, respectively.}
\end{deluxetable}

The detectability of the H$_2$O and O$_2$ features in the VIS bandpass varies greatly depending on the month in the simulation (Table \ref{tab:VIS_SN}). The maximum time required to detect the H$_2$O features in the 1.8 Gt case was 224 hours, whereas the maximum for the 3 other cases all exceed 900 hours. On the other hand, the minimum required observation time gets as low as 3 and 6 hours in the 15 Gt and 30 Gt cases, respectively. The minimum detection time for the 1.8 Gt and 60 Gt cases are slightly longer, where they require 30 and 37 hours, respectively. Detection of the O$_2$ feature in the VIS bandpass follows a similar trend where the feature can require extensive amounts of observation time to detect in some months, and be detected in as little as 18 hours in others. Similar to the H$_2$O features, the minimum time needed to detect the O$_2$ feature is shorter for the 15 Gt and 30 Gt cases than for the 1.8 Gt and 60 Gt cases.

\begin{deluxetable*}{|c|c|c|c|c|c|c|}
\tablecaption{Observation Hours Required to Detect VIS Features \label{tab:VIS_SN}}
\tablehead{\colhead{Eruption (Gt)} & \colhead{H$_2$O Max} & \colhead{H$_2$O Min} & \colhead{H$_2$O Avg} & \colhead{O$_2$ Max} & \colhead{O$_2$ Min} & \colhead{O$_2$ Avg}}
\startdata
1.8 & 224 & 30 & 107 & 75 & 26 & 37\\
15 & 934 & 3 & 184 & 1088 & 18 & 297\\
30 & 1321 & 6 & 185 & 1139 & 19 & 578\\
60 & 2939 & 37 & 232 & 1541 & 26 & 945\\
\enddata
\tablenotetext{}{Max, Min, and Med designate the maximum, minimum, and median time needed to detect the associated feature during each simulation, respectively.}
\end{deluxetable*}

The large fluctuation in the detectability of H$_2$O features is also apparent in the NIR bandpass (Table \ref{tab:NIR_SN}). The features are easiest to detect in the 15 Gt and 30 Gt eruption cases where they would require 9 and 10 hours of observation, respectively. Whereas the H$_2$O features require 29 hours in the 1.8 Gt case and 33 hours in the 60 Gt case. In all eruption cases, the H$_2$O features can require a maximum time for detection greater than 180 hours. Although the H$_2$O features in both the VIS and NIR can require extensive observation time in certain months, the VIS features have lower minimum observation times since they are at greater contrast ratios and therefore have greater signal.

It is interesting to note that the 1.8 Gt and 60 Gt cases have similar minimum detection times for the H$_2$O features in both the VIS and NIR bandpasses. In all the simulations, the force from the eruptions transport H$_2$O from the lower atmosphere into the upper atmosphere (Figure \ref{fig:TPs}). In the case of the 30 Gt eruption, H$_2$O vapor abundance in the stratosphere increased by 3 orders of magnitude \citep{guzewich2022volcanic}. This increase in H$_2$O vapor does not translate to larger H$_2$O features in the spectra during the first 4 years of the simulations however, because the influx of volcanic aerosols enhance scattering and reduce the size of the H$_2$O absorption features. Once the eruptions stop, the volcanic aerosol abundance begins to decrease which causes the H$_2$O absorption features to increase in depth over the last 4 years of the simulation (Figure \ref{fig:6monthplot}). The minimum detection time for the H$_2$O features in every eruption case are achieved during the final year of their simulation, which is when there is the least amount of volcanic aerosols. The larger abundance of volcanic aerosols produced by the 60 Gt eruption takes much longer to deplete than the other eruptions. By the end of the simulation there are still enough aerosols to significantly impact the size of the H$_2$O features, which increases their required detection time. The 1.8 Gt eruption produces far less volcanic aerosols than the other eruptions, which is why it has far less variation in feature detection time than the other eruptions (Tables \ref{tab:UV_SN}, \ref{tab:VIS_SN}, \& \ref{tab:NIR_SN}). But the 1.8 Gt eruption also has the smallest increase in the amount of H$_2$O vapor in the upper atmosphere. As a result, the longer minimum detection times for the H$_2$O features in the 1.8 Gt case are mainly due to lower H$_2$O vapor abundance in the upper atmosphere, rather than the presence of volcanic aerosols.


\begin{deluxetable}{|c|c|c|c|}
\tablewidth{\columnwidth}
\tabletypesize{\footnotesize}
\centering
\tablecaption{Observation Hours Required to Detect NIR Features \label{tab:NIR_SN}}
\tablehead{\colhead{Eruption (Gt)}  & \colhead{H$_2$O Max}  & \colhead{H$_2$O Min}  & \colhead{H$_2$O Avg}}
\startdata
1.8 & 182 & 45 & 89 \\
15 & 649 & 9 & 124 \\
30 & 982 & 10 & 129 \\
60 & 2546 & 42 & 172
\enddata
\tablenotetext{}{Max, Min, and Med designate the maximum, minimum, and median time needed to detect the associated feature during each simulation, respectively.}
\end{deluxetable}


\section{Discussion}
\label{sec:discussion}

\subsection{Inferring Volcanism from Reflectance Spectra}

Direct detection of SO$_2$ in the atmosphere of an exoEarth would provide strong evidence of persistent volcanic activity. This is because a consistent flux of SO$_2$ into the atmosphere would be required in order to sustain detectable amounts of SO$_2$, given its short lifetime in an Earth-like atmosphere. Absorption by SO$_2$ does occur at 0.29 $\mu$m, however it does not appear in the exoEarth spectra because it is overshadowed by O$_3$ and O$_2$ absorption (Figure \ref{fig:208411_Transmittance}). Detection of SO$_2$ is more plausible in a Venus-like atmosphere given its lack of O$_3$ and O$_2$. There are an abundance of potential exoVenuses that have been discovered \citep{ostberg2019, ostberg2023demographics}, however their vicinity to their host stars make them poor targets for direct imaging observations.

\begin{figure*}[htp]
  \includegraphics[width = \textwidth]{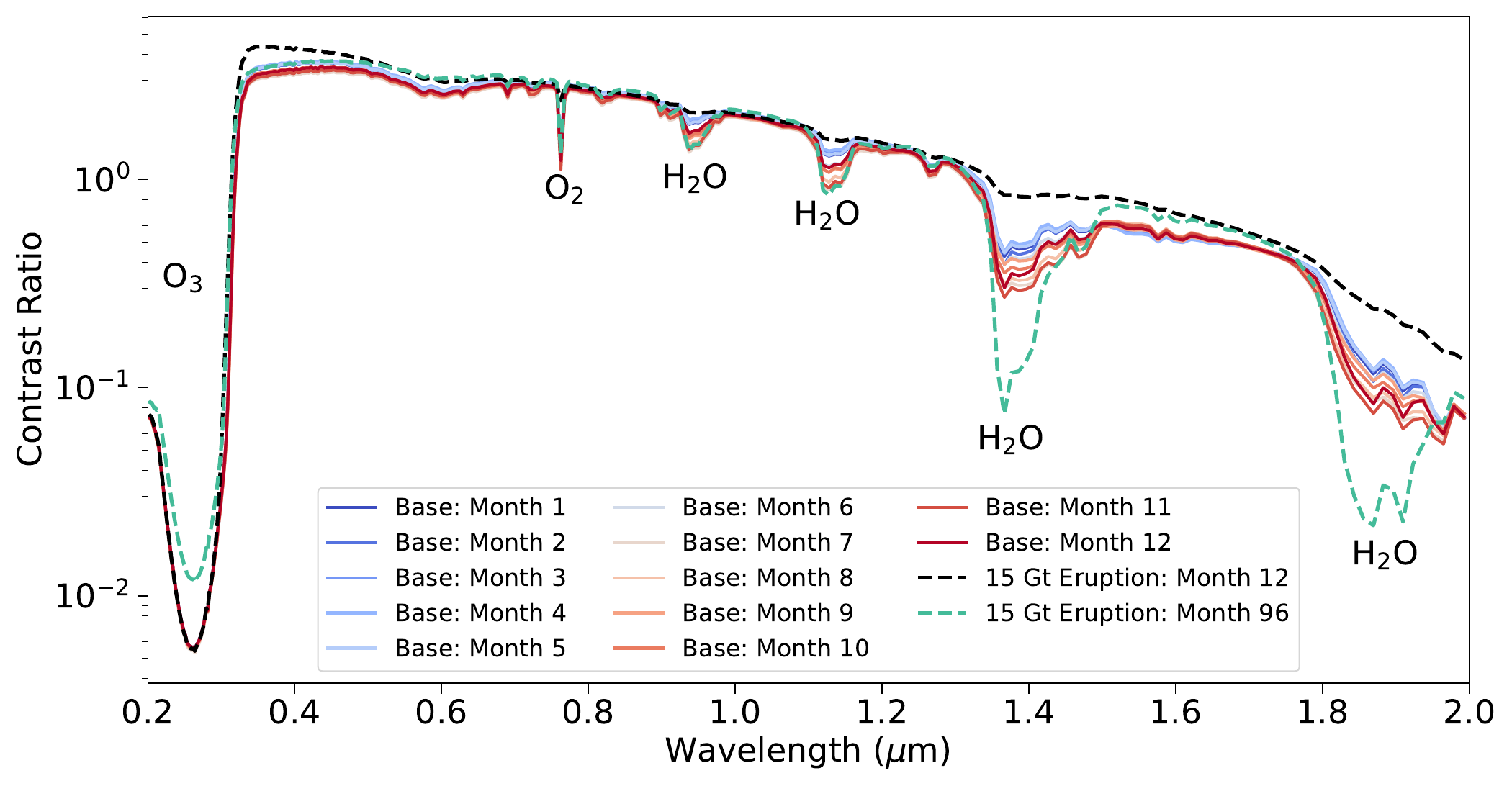}
  \includegraphics[width = \textwidth]{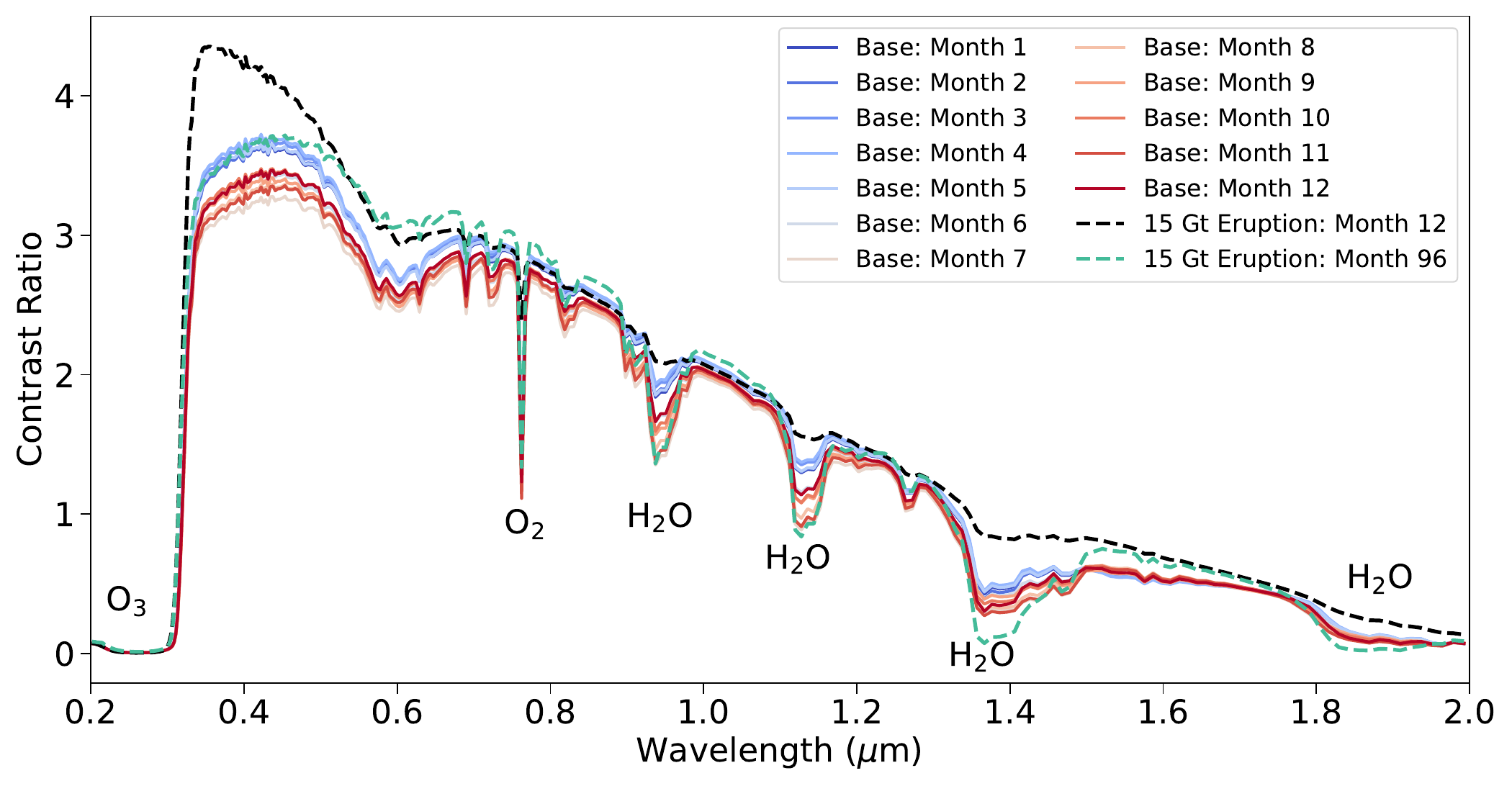}
  \caption{The reflectance spectra for each month of the baseline exoEarth simulation in comparison to months 12 and 96 of the 15 Gt eruption. Both plots show the same spectra but the upper panel is on a log scale and the lower panel is scaled linearly.}
  \label{fig:BaseSpectraPlots}
\end{figure*}

Since SO$_2$ absorption is not visible in the exoEarth spectra, volcanic activity would have to be indirectly inferred by detecting changes in the spectrum caused by volcanism. Atmospheric changes caused by seasonal effects can also lead to absorption features changing in size. To determine the magnitude of the spectral changes caused by seasonal effects, we modelled the reflectance spectrum of the baseline exoEarth simulation which does not include eruptions (Figure \ref{fig:BaseSpectraPlots}). The size of the O$_3$ and O$_2$ features remain essentially constant throughout the year in the baseline simulation, whereas both features have significant variation in the 15 Gt spectra. The H$_2$O features at 0.95 and 1.15 $\mu$m in the baseline spectra reach the same depth as those in the 15 Gt eruption in month 96, but the H$_2$O features at 1.35 and 1.85 $\mu$m reach a greater depth in the 15 Gt spectra than the baseline spectra. All H$_2$O features in the 15 Gt spectra in month 12 get much smaller than those in the baseline spectra, as they are almost entirely muted due to the effect of volcanic aerosols.

To avoid potential ambiguity when determining whether spectral changes are caused by seasonal effects or volcanic activity, features which have significantly more variability in the volcanic spectra than in the baseline spectra should be prioritized. To quantify variation in feature size we subtracted the minimum contrast ratio from the maximum contrast ratio of every feature in the 4 volcanic exoEarth cases and the baseline exoEarth case. The minimum value of a feature was defined to be the smallest contrast ratio achieved by a feature during throughout a given simulation, and the maximum value is the largest contrast ratio achieved by a feature. Both the maximum and minimum values were obtained at the peak wavelength of a given feature. Note that the minimum contrast ratio would be achieved when a feature reaches its largest depth.

Table \ref{tab:maxmin} lists the peak wavelength, and the maximum change in contrast ratio for the 6 major features in both the volcanic spectra (Volcanic Diff) and baseline spectra (Base Diff). We calculated the maximum change of a given feature in all 4 eruption cases, but only the largest change of the 4 cases was included in the table. The O$_3$ feature in the UV changed the most in the volcanic spectra during the 60 Gt eruption case, with a maximum change in contrast ratio of 0.0405 (Table \ref{tab:maxmin}). The same feature in the baseline spectra only has a change in contrast of 0.0002. In the 30 Gt eruption case the contrast ratio of the O$_2$ feature varied by 1.6138, which is the largest variation of the 6 major features. Unlike the O$_3$ feature, the O$_2$ feature has some variation in the baseline spectra with a maximum difference of 0.3499, but is still far less than that of the O$_2$ feature in the volcanic spectra (Table \ref{tab:maxmin}).

The variation of H$_2$O decreases towards longer wavelengths in both the volcanic and baseline spectra, where the H$_2$O feature at 0.95 $\mu$m has the most variation and the H$_2$O feature at 1.85 $\mu$m has the least variation (Table \ref{tab:maxmin}). In the volcanic spectra, the H$_2$O features change the most in the 15 Gt and 30 Gt eruption cases. The 1.8 Gt case has the lowest variation of all the eruption cases. The H$_2$O features at 0.95 and 1.15 $\mu$m in the 1.8 Gt spectrum have less variation than the same features in the base spectrum, and would be indiscernible from the variation caused by seasonal effects.

The O$_3$ feature in the UV is a potentially viable indicator of volcanism given that the variation of the feature is 3 orders of magnitude greater in the volcanic spectra than in the baseline spectra (Table \ref{tab:maxmin}). The magnitude of the variation is the smallest of all features however, meaning that detecting the change in size of the feature would require more sensitive instrumentation than the other features. The O$_2$ feature has the largest magnitude of variation, and the largest difference in variation when comparing the variation of the feature in the volcanic spectra to the baseline spectra. The main downside of the O$_2$ feature is that it is the thinnest of all the features and would require high spectral resolution to detect. The difference in maximum variation between the volcanic and baseline spectra is roughly the same for all H$_2$O features. However, the H$_2$O feature in the VIS bandpass at 0.95 $\mu$m should be prioritized over the other H$_2$O features since it has the largest magnitude of variation (Table \ref{tab:maxmin}) and has the best minimum detection time (Table \ref{tab:VIS_SN}).

The sharp peak which forms around 0.4 $\mu$m could also be an indicator of volcanism (Figures \ref{fig:6monthplot} \& \ref{fig:BaseSpectraPlots}). The peak appears only in the volcanic spectra and never in the baseline spectra, which removes any potential ambiguity involved with discerning seasonal and volcanic effects. Unlike using features to infer volcanism which requires detecting variation across multiple observations, detection of the peak alone would provide evidence of volcanism and could potentially be done in a single observation. A Venus-like planet that can sustain SO$_2$ in its atmosphere for an extended period of time could be a potential false-positive since it may yield a similar peak in its spectrum.

\begin{deluxetable}{ c c c c }
\tabletypesize{\scriptsize}
\tablecaption{Feature Contrast Ratio Variation}
\tablehead{\colhead{Feature} & \colhead{Peak Wavelength} & \colhead{Volcanic Diff} & \colhead{Baseline Diff} }
\startdata
O$_3$ & 0.25 & 0.0405 & 0.0002 \\
O$_2$ & 0.75 & 1.6138 & 0.3499 \\
H$_2$O & 0.95 & 1.3948 & 0.5813 \\
H$_2$O & 1.15 & 1.1411 & 0.4857 \\
H$_2$O & 1.35 & 0.8064 & 0.1795 \\
H$_2$O & 1.85 & 0.1815 & 0.0373 \\
\enddata
\tablecomments{The maximum change in contrast ratio for major features in the 4 volcanic exoEarth spectra (Volcanic Diff), and the baseline exoEarth spectra with no eruptions (Baseline Diff).}
\label{tab:maxmin}
\end{deluxetable}


\subsection{Spectral Dependence on Phase Angle and Observed Longitude}

All reflectance spectra were modelled assuming the illuminated region of the planet facing the observer was latitude 0$^{\circ}$ and longitudes 180$^{\circ}$--270$^{\circ}$, which consists mostly of the Americas and the Pacific Ocean. Changing the illuminated region to include only the Pacific Ocean would likely decrease the average contrast ratio of the spectra since water is less reflective than land. We expect the molecular absorption to be relatively constant between regions since we used monthly averaged atmospheres to model the reflectance spectra. 

If we were to compare regional reflectance spectra on shorter timescales, then we would expect there to be spectral differences because of localized weather patterns and cloud coverage. In addition, the cadence of observations could cause the phase and illuminated region of the planet to differ in each observation. These variations in weather and viewing geometry can lead to discrepancies in the reflectance spectra from separate observations. Determining if these spectral variations are distinguishable from those caused by volcanism should be investigate in future studies to confirm whether changes in absorption features can be a reliable indicator of volcanic activity.


\subsection{Likelihood of Observing an ExoEarth with Ongoing Volcanism}

Detecting signs of volcanic activity in an exoplanet atmosphere requires observations to be conducted while volcanism is occurring on the planet, or while the atmosphere still has remnants of past volcanism. The frequency of LIP volcanism on Earth has varied throughout time. From present day to 180 Mya, LIPs occurred about once every 10 My \citep{coffin2001large,ernst2005frontiers}, but from 180 to 2600 Mya, LIPs occurred once every 20 My \citep{ernst2014large}. The duration that a LIP remains active can be as long as tens of millions of years, but most cases of LIPs are more brief and last only 0.5--1.0 My \citep{hofmann200040ar,courtillot2003ages,jerram2005anatomy,blackburn2013zircon,ernst2014large}. 

We assumed that on average, LIPs on Earth have occurred every 15 million years and lasted for 1 million years. Which means if you were to choose a random year over the last 3 Gy of Earth's history, there is roughly a 6.6\% chance of there being an active LIP. If we assume that exoEarths targeted in future direct imaging missions also have a 6.6\% chance of having an active LIP, then at least 47 planets would need to be observed to have greater than a 90\% chance that at least one planet has ongoing volcanism. This is an optimistic, first-order approximation since it is contingent on a variety of assumptions, but demonstrates that a significant amount of observing time will be required to have an opportunity to detect volcanism on an exoplanet.


\section{Conclusions}
\label{sec:conclusions}

In this work we explored the possibility of detecting volcanism in the reflectance spectrum of an exact exoEarth analog orbiting a Sun-like star at 1 AU. The primary absorption features in the reflectance spectra were O$_3$ in the UV bandpass, O$_2$ and H$_2$O in the VIS bandpass, and H$_2$O in the NIR bandpass. We determined the range of detectability for each absorption feature in all reflectance spectra for all 4 eruption cases. Absorption by O$_3$ was both the easiest to detect and most consistently detectable feature of the group since absorption was always present in every spectra. The detectability of every other feature varied greatly on a monthly basis. In particular, the H$_2$O absorption features were almost entirely concealed by volcanic aerosols while eruptions were ongoing, and continuously grew in size once the eruptions ceased and the volcanic aerosols began to be removed from the atmosphere.

Detecting SO$_2$ in the atmosphere of an exoEarth would provide strong evidence for ongoing volcanism given the short lifetime of SO$_2$ in Earth's atmosphere. While eruptions were ongoing in the simulations, absorption by SO$_2$ contributed to the reflectance spectra, but was always hidden beneath the stronger O$_3$ absorption feature which appears around 0.3~$\mu$m. Since SO$_2$ absorption is undetectable in the simulated exoEarth spectra, we propose the best method for inferring volcanism is through detecting spectral changes caused by volcanism, or by observing the sharp peak which forms between 0.3--0.5 $\mu$m.

We quantified the maximum amount of variation incurred by each of the major features in the reflectance spectra of each of the volcanic exoEarths and the baseline exoEarth. The maximum variation of every feature in the volcanic spectra was greater than the variation of the same features in the baseline spectra. The 1.8 Gt eruption spectra was the only eruption case which had features with less variation than the baseline spectra, which were the H$_2$O features at 0.95 and 1.15 $\mu$m. The O$_3$ feature, and O$_2$ and H$_2$O features in the VIS bandpass have the largest discrepancy in variation between the volcanic and baseline spectra, making them optimal features for discerning spectral changes caused by volcanism from those caused by seasonal effects. The sharp peak which forms in the volcanic spectra between 0.3--0.5 $\mu$m is also a potentially strong indicator of ongoing volcanism since it only formed during times of high volcanic aerosol abundance.

Future work will be required to investigate whether short term changes in weather or cloud coverage may cause fluctuations in spectral features similar to those caused by volcanism. If changes in spectra features from weather and volcanism are similar, then changes in feature size alone may not be a reliable indicator of ongoing volcanism. This work considered a small wavelength range, but other wavelength ranges should be investigated to see if other volcanic indicators, such as absorption by SO$_2$, could be detected in other bandpasses. Eruptions of different compositions should be tested as well to determine if some eruptions may be easier to detect than others. Modelling eruptions similar to the Hunga Tonga eruption, which delivered massive amounts of H$_2$O vapor into the atmosphere, may have a far more significant effect on H$_2$O absorption features and could prove to be a more reliable signal then what was discussed in this work.

Direct imaging missions will be our first opportunity to characterize the atmospheres of exoEarths around Sun-like stars. These missions are planned to launch in at least the next decade, but in the meantime it is vital that we refine our ability to analyze such data so that we may maximize what can be learned from it. In particular, learning how to identify possible indicators of volcanism on an exoplanet will be crucial since it can provide invaluable insight into the state of the planet's interior which would otherwise be inaccessible to us. Improving our understanding of volcanic outgassing within the solar system is also important, so that we may better infer the likelihood of volcanism on exoplanets \citep{horner2020b,kane2021d}. These parallel data sources will help improve understanding of the different evolutionary pathways of terrestrial planets, and potentially identify which planets may have surface conditions suitable for life.


\section*{Acknowledgements}
Ostberg and Kohler were supported by the NASA Deep Atmosphere Venus Investigation of Noble gases, Chemistry, and Imaging (DAVINCI) mission. Guzewich, Oman, Fauchez, Kopparapu, Richardson, and Whelley were supported by NASA GSFC's Sellers Exoplanet Environments Collaboration. Kane was supported by NASA grant 80NSSC21K1797, funded through the NASA Habitable Worlds Program. GEOSCCM is supported by the NASA MAP program and the high-performance computing resources were provided by the NASA Center for Climate Simulation (NCCS). The material is based upon work supported by NASA under award number 80GSFC21M0002.

\bibliographystyle{aasjournal}
\bibliography{references,nea}


\end{document}